\documentclass[aps,prd,twocolumn,showkeys,floats,showpacs, superscriptaddress,nofootinbib]{revtex4-1}
\usepackage{graphicx}
\usepackage{epsfig}
\usepackage{latexsym}
\usepackage{amssymb}
\usepackage{bm}
\usepackage{float}
\usepackage{subfigure}
\usepackage{amsmath}
\usepackage{amsfonts}
\usepackage{color}

\usepackage[colorlinks=true]{hyperref}

\usepackage{aas_macros}

\def\L {\mathcal{L}}

\def\g{\sqrt{-g}}

\def\be{\begin{equation}}
\def\ee{\end{equation}}
\def\bea{\begin{eqnarray}}
\def\eea{\end{eqnarray}}

\begin{document}

\title{Rethinking the link between matter and geometry}

\date{\today}
\author{Olivier Minazzoli}
\affiliation{Centre Scientifique de Monaco, 8 Quai Antoine 1er, Monaco}
\affiliation{Laboratoire Artemis, Universit\'e C\^ote d'Azur, CNRS, Observatoire C\^ote d'Azur, BP4229, 06304, Nice Cedex 4, France}
\email{ominazzoli@gmail.com}

\begin{abstract}

In the present manuscript, I examine an intriguing relation at the classical level between general relativity and a theory where matter couples uniquely multiplicatively to geometry in the Lagrangian density. Interestingly, the gravitational constant $G$ is replaced by a novel fundamental constant, whose value is not tied to any classical phenomenon; while the value of $G$ itself becomes related to the dynamics of the universe. I concentrate on different aspects of the Equivalence Principle, as the theory is expected to violate all of its different formulations.
\end{abstract}

\maketitle

\section{Introduction}


In most attempts to find a new theory of gravity, parts of the Lagrangian that correspond to matter are usually considered separately from the geometric one, before being summed. In other words, one assumes that the two sectors of nature can be described separately before being put together in order to consider them simultaneously. 

From a historical perspective, it is a rather legitimate assumption, given the fact that a good description of matter fields has originally been (and still mostly is) formulated in the context of special relativity --- for which gravity is entirely absent. Hence, from this `experience', it seems legitimate to assume that at the level of the action, matter and gravity can be described by distinct Lagrangians, which can then be summed in order to describe both of the two worlds at the same time. 

Besides, beyond its apparent legitimacy, this assumption also played a central role in order to find general relativity --- which, perhaps more than anything else, is a very satisfactory (\textit{a posteriori}) justification of the Equivalence Principle that continues to be respected to always higher accuracy, e.g. \cite{touboul:2017pr,viswanathan:2018mn,safronova:2018rm}.  

Nevertheless, one may question whether assuming that matter fields and gravity are described by separate Lagrangians is a good guiding principle in order to search for a better picture of fundamental physics than the current accepted model. For instance, in 1918, Einstein stated in a letter to Weyl \cite{pais:1982bk}: \textit{``Ultimately it must turn out that action densities must not be glued together additively. I too, concocted various things, but time and again I sank my head in resignation.''}

In particular, it is tantalizing to imagine a greater interconnection between matter fields and geometry in the context of a general quantum theory of fundamental physics.

A surprising theory that goes beyond this assumption has recently been suggested \cite{ludwig:2015pl}. It is based on a $f(R,\L_m)$ action in four dimensions \cite{harko:2010ep}. As we shall see, (at least) part of its phenomenology may nevertheless be very similar to the one of general relativity in some of its (classical) limits. 

\section{The theory}

The action of the `surprising theory' reads as follows:
\be
S=-\frac{\xi}{2c} \int d^4x \sqrt{-g} \frac{\L_m^2}{R}, \label{eq:actionfR}
\ee
as opposed to the one of general relativity, which reads
\be
S=\frac{1}{c} \int d^4x \sqrt{-g} \left(\frac{R}{2 \kappa}+\L_m \right), \label{eq:actionGR}
\ee
with $\kappa \equiv 8\pi G/c^4$, $c$ and $G$ are the speed of light and gravitational constant respectively, $R$ is the Ricci scalar, while $\L_m$ represents the Lagrangian of matter fields and $\xi$ is a constant with the dimension of $\kappa$. \footnote{Note that, for dimensional reasons, $\xi$ is a numerator while $\kappa$ is a denominator, even though they have the same dimension.} The most important thing to notice right away is that in this framework, it no longer makes sense to consider matter without geometry, and vice versa. In particular, according to Eq. (\ref{eq:actionfR}), one may expect any space-time that gives $R=0$ (including Minkowski's space-time) not to be an exact solution of this theory. The metric field equation derived from (\ref{eq:actionfR}) reads
\be
R_{\mu \nu}-\frac{1}{2} g_{\mu \nu} R= - \frac{R}{\L_m} T_{\mu \nu} + \frac{R^2}{\L_m^2} \left(\nabla_\mu \nabla_\nu - g_{\mu \nu} \Box \right)\frac{\L_m^2}{R^2}, \label{eq:fRmetricfield}
\ee
with the usual stress-energy tensor definition:
\be
T_{\mu \nu} \equiv - \frac{2}{\sqrt{-g}} \frac{\delta \left(\sqrt{-g} \L_m \right)}{\delta g^{\mu \nu}}. \label{eq:defSET}
\ee
One important thing to notice is that the constant $\xi$ does not appear in the classical equations of motion. Therefore, this constant can only be relevant at the quantum level --- through, for instance, the path functional integration of action (\ref{eq:actionfR}). 

In what follows, I concentrate on the classical side of the theory. We will see that the trace of the metric field equation is particularly illuminating. It reads
\be
3 \frac{R^2}{\L_m^2} \Box \frac{\L_m^2}{R^2}=  R \left(1 -\frac{T}{\L_m}\right). \label{eq:tracemetricFE}
\ee

\section{Building some intuition}

One may ask, however, how such a theory may have any similarity with general relativity. A first hint to this question comes from assuming that the on-shell Lagrangian can be equal to the trace of the stress-energy tensor --- at least in some specific situations. By that, I mean that I assume that there may be solutions to the matter field equations that are such that $\L_m = T$ when the actual solutions of the fields are injected in the formal equation of $\L_m$. In those situations, the right-hand side of Eq. (\ref{eq:tracemetricFE}) vanishes, such that $\L_m/R=$ constant is a solution of Eq. (\ref{eq:tracemetricFE}). For this solution, Eq. (\ref{eq:fRmetricfield}) reduces to the equation of general relativity:
\be
R_{\mu \nu}-\frac{1}{2} g_{\mu \nu} R=\frac{8 \pi G}{c^4}T_{\mu \nu}, \label{eq:Einstein}
\ee
with the following identification:
\be
 \frac{R}{\L_m}=- \frac{8 \pi G}{c^4}, \label{eq:dustlimitG}
\ee
which --- since $\L_m = T$ --- turns out to be nothing else than the trace of the Einstein equation $R = - \frac{8 \pi G}{c^4} T$. This would mean that in this context, the constant of Newton is not a fundamental constant of the theory, but emerges as a (specific) solution of the dynamics of the field equations \footnote{i.e. $G \equiv \frac{c^4 R}{8\pi \L_m} \sim$ constant.}. On the other hand, whenever $\L_m \neq T$, Eq. (\ref{eq:tracemetricFE}) drives $R/\L_m$ away from a constant, and therefore, Eq. (\ref{eq:fRmetricfield}) drives the theory away from general relativity. 

In general relativity already, one may expect both situations ($\L_m = T$ and $\L_m \neq T$) to occur, depending on the underlying physics. \footnote{Of course, for general relativity, this discussion would be pointless given the fact that the Lagrangian does not explicitly appear in the field equations.} For instance, take an electric field in a given direction, the on-shell Lagrangian reduces to the energy density, while the electromagnetic tensor is traceless (hence $\L_m \neq T$). On the other side, for an electromagnetic radiation, given that the modulus of the electric and magnetic fields are equal, the on-shell Lagrangian vanishes, just as the trace of the electromagnetic stress-energy tensor does (hence $\L_m = T$). For a dust field on the other side, one deduces from $S = -mc^2 \int d\tau$ that the Lagrangian is proportional to the rest-mass energy density (hence $\L_m = T$). Etc.

However, the relation with matter fields in the present theory becomes more involved because the matter field equations are also modified with respect to general relativity, as soon as $R/\L_m \neq$ constant. Indeed, for any tensorial field $\chi$, the Euler-Lagrange equation is modified according to
\be
\frac{\partial \L_m}{\partial \chi} -  \frac{1}{\g} \partial_\sigma \left(\frac{\partial \g \L_m}{\partial (\partial_\sigma \chi)}  \right) = \frac{\partial \L_m}{\partial (\partial_\sigma \chi)} \partial_\sigma \ln \left(\frac{\L_m}{R} \right).\label{eq:matter_fRL}
\ee
Of course, the conservation equation is in general modified as well, such that one has
\be
\nabla_\sigma \left(\frac{\L_m}{R} ~T^{\alpha \sigma} \right) = \L_m \nabla^\alpha\left(\frac{\L_m}{R}\right), \label{eq:cons_fRL}
\ee
almost necessarily leading to a violation of the Equivalence Principle as soon as $\L_m / R \neq$ constant \footnote{This remark will become clearer with Eq. (\ref{eq:actionscalar}).}. But again, if one assumes that Eq. (\ref{eq:dustlimitG}) is possible in some limit --- given that $\L_m$ may be equal to $T$ in some limit --- then it is possible that the theory becomes classically equivalent to general relativity in this limit.

It may be worth noticing that in the general context of $f(R,\L_m)$ theories, it has been argued that the average on-shell Lagrangian of localized concentrations of energy with fixed rest mass and structure (solitons) is the trace of the stress-energy tensor \cite{avelino:2018pr,*avelino:2018pd}. If correct, it gives an example for which the theory in Eq. (\ref{eq:actionfR}) can reduce to general relativity --- perhaps independently of the underlying specifics of the matter sector.
\section{An almost equivalent theory}

A convenient way to understand the classical equivalence with general relativity in the limit such that $\L_m = T$ is to use the scalar-tensor (almost) equivalent action to Eq. (\ref{eq:actionfR}), which reads \cite{ludwig:2015pl}
\be
S=\frac{1}{c}\frac{\xi }{\kappa} \int  d^4x \sqrt{-g} \left[\frac{\phi R}{2 \kappa}+ \sqrt{\phi} \L_m \right]. \label{eq:actionscalar}
\ee
Indeed, one can show that it classically corresponds to Eq. (\ref{eq:actionfR}) with the identification $\sqrt{\phi} = - \kappa \L_m/R$ \cite{ludwig:2015pl}. \footnote{Obviously, the correspondence breaks at $R=0$.}
Equation (\ref{eq:actionscalar}) leads to the following equations of motion
\be
G_{\alpha \beta} = \kappa \frac{T_{\alpha \beta}}{\sqrt{\phi}} + \frac{1}{\phi} \left[\nabla_\alpha \nabla_\beta - g_{\alpha \beta} \Box \right] \phi, \label{eq:stt_metric}
\ee
where $G_{\alpha \beta}$ is the Einstein tensor, as well as
\be
\frac{3}{\phi} \Box \phi =\frac{ \kappa}{\sqrt{\phi}} \left(T-\L_m\right),
\ee
and the conservation equation reads
\be
\nabla_\sigma \left(\sqrt{\phi} T^{\alpha \sigma} \right) = \L_m \nabla^\alpha \sqrt{\phi}.\label{eq:stt_conserv}
\ee
At the same time, the matter field equations are modified with respect to the case where general relativity is minimally coupled to matter. Indeed, one has
\be
\frac{\partial  \L_m}{\partial \chi} - \frac{1}{\g} \partial_\sigma \left(\frac{\partial \g \L_m}{\partial (\partial_\sigma \chi)} \right) = \frac{\partial \L_m}{\partial (\partial_\sigma \chi)} \partial_\sigma \ln \sqrt{\phi}.
\ee
See a specific case for instance in \cite{hees:2014pr}. The correspondence with Eqs. (\ref{eq:fRmetricfield}), (\ref{eq:tracemetricFE}), (\ref{eq:matter_fRL}) and (\ref{eq:cons_fRL}) is obvious. Also, if $\L_m = T$, $\phi = \phi_0=$ constant is a solution of the scalar-field equation. In that situation, one recovers the equations of general relativity minimally coupled to matter fields, as Eqs. (\ref{eq:stt_metric}) and (\ref{eq:stt_conserv}) then reduce respectively to 
\be
G_{\alpha \beta} = \kappa \frac{T_{\alpha \beta}}{\sqrt{\phi_0}}, \textrm{ and }\nabla_\sigma T^{\alpha \sigma}  =0,
\ee
while the matter field equations also reduce to the equations that are derived in general relativity.

On the other side, whenever $\L_m \neq T$, then the scalar-field cannot be a constant, such that it should lead to a violation of the universality of free fall \cite{damour:2010pr, minazzoli:2016pr} and to various violations of the Equivalence Principle (e.g. at the cosmological level \cite{hees:2014pr}) that are severely constrained by observations. Therefore, one has to check whether or not the actual scalar-field solution can be close enough to be a constant that the theory can pass Equivalence Principle tests. 

\section{Cosmological phenomenology and the Equivalence Principle}

It can easily be checked that if one assumes a flat Friedmann-Lema\^itre-Robertson-Walker (FLRW) metric for the universe, the scalar field is indeed quickly driven toward a constant --- at least during the matter era \cite{minazzoli:2014pr,minazzoli:2014pl}. However, the theory described by Eq. (\ref{eq:actionscalar}) alone cannot account for the acceleration of the expansion of the universe, because it converges toward general relativity without a cosmological constant --- as one can check from \cite{minazzoli:2014pr}. Therefore, one may have to consider quantum corrections to Eq. (\ref{eq:actionfR}) in order to get a theory that seems consistent with cosmological observations. 

\subsection{The cosmological constant}

Still, note that it has been shown that a quadratic potential in addition to the Lagrangian given in Eq. (\ref{eq:actionscalar}) can lead to an acceleration of the expansion of the universe, without spoiling the constancy of the scalar-field at the cosmological level \cite{minazzoli:2014pr}, and therefore without spoiling the satisfaction of the Equivalence Principle at the cosmological level \cite{hees:2014pr}. However, unless quantum field theory applied on Eq. (\ref{eq:actionfR}) can generate corrections such as a quadratic potential when the theory is written in its scalar-tensor form Eq. (\ref{eq:actionscalar}), it seems unlikely that the theory given in Eq. (\ref{eq:actionfR}) can be consistent with cosmological observations. Nevertheless, one can see that if such a `corrected theory' had to pass cosmological tests, it could read as follows
\be
S=-\frac{\xi}{2c} \int d^4x \sqrt{-g} \left(\frac{\L_m^2}{R} + \alpha \frac{\L_m^4}{R^4}\right), \label{eq:actionfR_lambda}
\ee
where $\alpha$ is a constant with the dimension such that $\alpha/\kappa^2 \equiv 2\Lambda$ has the dimension of an inverse squared length.
 Indeed, regarding the extra term as a perturbation and using the zeroth order identification $\sqrt{\phi} = - \kappa \L_m/R$, this action reduces to
\be
S=\frac{1}{c}\frac{\xi }{\kappa} \int  d^4x \sqrt{-g} \left[\frac{1}{2 \kappa} \left(\phi R-2\Lambda \phi^2 \right)+ \sqrt{\phi} \L_m \right]. \label{eq:actionscalar_lambda}
\ee
It is important to stress, however, that if the additional term in Eq. (\ref{eq:actionfR_lambda}) is not a perturbation of the original theory --- but rather is a part of the zeroth order classical field equations --- then Eq. (\ref{eq:actionfR_lambda}) does not correspond to Eq. (\ref{eq:actionscalar_lambda}), because the variation of $\phi$ with respect to the metric would no longer cancel for all the terms in the action. It is therefore a rather strong restriction, which may turn out to be a shortcoming of the theory. (Note, however, that Eq. (\ref{eq:actionscalar_lambda}) can be considered as a stand-alone theory --- although it would be much less exotic and intriguing than the $f(R,\L_m)$ theory I started with). But the cosmological (flat FLRW) solution of the theory described by Eq. (\ref{eq:actionscalar_lambda}) exponentially converges toward general relativity with a cosmological constant $\Lambda$ during the matter and dark energy eras --- as can be seen in section III.E. of \cite{minazzoli:2014pr} --- without spoiling the Equivalence Principle in cosmological observables since the scalar-field is exponentially driven toward a constant (at least during the matter and dark energy eras) --- as one can infer from \cite{hees:2014pr} and \cite{minazzoli:2014pr}.

\subsection{Dynamical coupling constants}

What the convergence toward general relativity would also mean is that the Newtonian constant today 
\be
G_\textrm{today} \equiv \left. \frac{c^4 R}{8\pi \L_m}\right|_{z=0}, 
\ee
where $z$ is the cosmological redshift, is the result of the cosmological evolution of the field equations Eq. (\ref{eq:fRmetricfield}). Besides, since the scalar-field degree of freedom (nonminimally) couples to matter in Eq. (\ref{eq:actionscalar_lambda}), the same goes for other coupling constants \cite{damour:2010pr,minazzoli:2016pr} --- such as the fine structure constant \cite{hees:2014pr}. In the end, one would have replaced \textit{absolute (or `rigid') structures} (the coupling constants) by \textit{dynamical (or `elastic') entities} (see \cite{damour:2012cq} for a discussion on these aspects). It is rather interesting to remark that another approach that initially aimed at merging matter and geometry in the same Lagrangian --- the Kaluza-Klein theory --- also led to a relaxation of (part of the) \textit{absolute structures} \cite{damour:2012cq}. In other words, the theory presented in Eq. (\ref{eq:actionfR}) is a way to merge matter and geometry in a unique Lagrangian that is an alternative to the (old) Kaluza-Klein idea of adding dimensions. But it is important to stress how little room seems to be left for modifying the function $f(R,\L_m)$ in Eq. (\ref{eq:actionfR}), if one wants the following three properties to be satisfied: 1/ matter and geometry are not separable in the action, 2/ the dynamics should tend toward general relativity in some (observable) limits and 3/ the Equivalence Principle should not be `strongly' violated in (observable) situations. Indeed, the \textit{intrinsic decoupling} that leads to the (potential) good behavior of the theory comes from an exact cancellation of different terms in the field equations. Hence, it does not seem likely that any other $f(R,\L_m)$ theory can satisfy the three premises stated above.

In some sense, in order to get the action Eq. (\ref{eq:actionfR}) --- instead of the action Eq. (\ref{eq:actionGR}) --- one has to relax the Equivalence Principle as a fundamental guiding principle \footnote{Note that it has already been argued that the Equivalence Principle should not be counted among the basic principles of physics, see e.g. \cite{damour:2012cq}.}, and to use another guiding principle that demands geometry to be tied to matter in an inseparable way. Then, the fact that the theory must explain observations seems to fix the remaining freedom in choosing the action.

\subsection{Another perspective: A conformal representation}

The relation of the theory described by Eq. (\ref{eq:actionscalar_lambda}) with general relativity with a cosmological constant is more direct in a conformal representation such that $g^*_{\alpha \beta} = \phi g_{\alpha \beta}$. Indeed, in this conformal representation, the kinetic gravitational term takes the form of general relativity with a cosmological constant $\Lambda$ and a scalar-field decoupled from the Ricci scalar:
\bea
 \sqrt{-g} \frac{1}{2 \kappa}&& \left(\phi R -2\Lambda \phi^2\right) \label{eq:conf_trans}\\
\nonumber &&=\sqrt{-g^*} \frac{1}{2 \kappa}\left(  R^* -2\Lambda  - \frac{1}{2} g_*^{\alpha \beta} \partial_\alpha \varphi \partial_\beta \varphi \right),
\eea
with $\varphi \equiv \sqrt{3} ~\ln \phi ~+$ arbitrary constant --- up to a total derivative that does not contribute to the equations of motion. The asterisk indicates that a quantity is defined with the conformal metric $g^*_{\alpha \beta}$. At the same time, the scalar-field equation is only sourced by  a term proportional to $\L_m^* - T^*$, \footnote{As can be inferred from Eq. (10) in \cite{minazzoli:2014pl}.} which is (close to be) null for a nonrelativistic matter content --- even with the cosmological term in Eq. (\ref{eq:conf_trans}). As a consequence, the cosmological friction term related to the Hubble-Lema\^itre function $H=\dot a /a$ quickly freezes the scalar field to its local value during the evolution of the universe --- at least after the radiation era. Hence, it gives another perspective as to why the theory described by Eq. (\ref{eq:actionscalar_lambda}) indeed cosmologically converges toward general relativity with a cosmological constant (henceforth satisfying the Equivalence Principle to some degree) --- at least after the radiation era. 

\subsection{Potential new prospects for early-universe models}

On the other side, since the scalar-field equation is sourced by a term proportional to $\L_m^* - T^*$, if the universe had happened to be dominated by a scalar-field during its infancy --- say, by the kinetic energy of the infant Higgs field --- then one can expect that it would have led to a different dynamics than for general relativity with a cosmological constant and a scalar field. Indeed, one has $\L_m^* - T^* \neq 0$ for scalar fields. Therefore, one may use this property in order to build new early-universe models. 

\section{The universality of free fall}

Inferring whether or not the theory described by Eq. (\ref{eq:actionfR}) respects the universality of free fall --- at least to a given level accuracy --- may also be extremely complicated. Indeed, in the current picture of fundamental physics --- general relativity with a cosmological constant plus the standard model of particles --- most of the mass of atoms comes from quantum (trace) anomalies of matter fields \cite{donoghue:1992bk}. Hence, it seems that one has no choice but to investigate the quantum side of the theory described by Eq. (\ref{eq:actionfR}) in order to figure out whether or not it respects the universality of free fall, at least to a given level of accuracy. Nevertheless, in the framework of a scalar-tensor theory with nonminimal couplings to matter --- such as the theory described by Eq. (\ref{eq:actionscalar}) or (\ref{eq:actionscalar_lambda}) --- it has been shown in \cite{minazzoli:2016pr} that if the scalar field universally couples to the trace of the `quantum corrected' matter stress-energy tensor \footnote{See also \cite{nitti:2012pr} for a discussion on the relevance of also taking into account the QED trace anomaly, while in \cite{minazzoli:2016pr}, Hees and I only considered the QCD trace anomaly --- following the seminal work of Damour and Donoghue \cite{damour:2010pr}. Note, however, that the splitting between purely QCD and QED effects is ambiguous when QED is also turned on \cite{gasser:2003ej}.}, and if this coupling is the square root of the coupling with the Ricci scalar --- which is the case in Eqs. (\ref{eq:actionscalar}) and (\ref{eq:actionscalar_lambda}) --- then the scalar field actually decouples from the matter dynamics, and one recovers the universality of free fall as well as the rest of the phenomenology of general relativity. Therefore, it seems plausible that the theory given by Eq. (\ref{eq:actionfR}) --- or by Eq. (\ref{eq:actionscalar_lambda}) --- respects the universality of free fall --- at least to a certain level of accuracy --- even though it may be very demanding to actually evaluate it.

\section{Conclusion}

In conclusion, the theory presented in this manuscript presents several interesting features:
\begin{itemize}
\item It is simple in its formulation, and yet (very) exotic. In particular, it does not require any new field. Furthermore, at the classical level, it has one parameter less than general relativity, because the Newtonian constant $G$ becomes a dynamical entity, while the coupling constant $\xi$ (replacing $\kappa=8\pi G/c^4$ in the action Eq. (\ref{eq:actionfR})) can be relevant at the quantum level only.
\item It has a richer dynamics than general relativity, but its classical dynamics tends toward the one of general relativity during the expansion of the universe. Notably, features usually expected from the Equivalence Principle (e.g. the universality of free fall) would no longer be fundamental in this theory, but emergent.
\item It is a new way to `entangle' matter and geometry in a single Lagrangian --- which is, otherwise, often achieved by adding dimensions.
\item It seems that it cannot be `deformed' if one wants to recover general relativity and the Equivalence Principle in some limits.
\item The Planck units are no longer tied to a fundamental gravitational parameter in this theory. However, new fundamental units could in principle be defined from the novel coupling constant $\xi$ in Eq. (\ref{eq:actionfR}). Moreover, given the fact that $\xi$ does not appear in the classical field equations, it is tempting to conjecture that this parameter may indeed be related to a quantum field theory that encompasses both the matter and the gravitational sectors.
\end{itemize}
On the other hand, one has to investigate the complicated quantum field theory side of the theory in order to quantitatively estimate its viability --- in particular with respect to both the acceleration of the expansion of the universe and the universality of free fall issues.
However, if this theory has any truth in it, it seems very likely that the matter sector will also have to be modified with respect to the current `standard' --- which was designed in the context of special relativity.

Finally, because it cannot be excluded \textit{a priori} that nature has several ways that entangle matter and geometry at play simultaneously, let us note that it may also be interesting to see how this particular theory behaves in more than 4 dimensions, as well as to check whether or not there exist alternative functions $f(R,\L_m)$ that possess a similar `intrinsic decoupling' in 4 --- or more than 4 --- dimensions.

\end{document}